\documentclass{article}
\usepackage{spconf,graphicx}
\usepackage{xcolor}
\usepackage{lipsum}
\usepackage{hyperref}
\usepackage{makecell}
\usepackage{enumitem}
\usepackage{comment}
\usepackage{mwe}
\setlist{nosep, leftmargin=14pt}

\title{Gray matter segmentation in ultra high resolution 7 Tesla ex vivo T2w MRI of human brain hemispheres}
\name{\begin{tabular}{c}Pulkit Khandelwal$^{1}$, Shokufeh Sadaghiani$^{1}$, Michael Tran Duong$^{1}$, Sadhana Ravikumar$^{1}$, Sydney Lim$^{1}$, \\
\textit{Sanaz Arezoumandan$^{1}$, Claire Peterson$^{1}$, Eunice Chung$^{1}$, Madigan Bedard$^{1}$, Noah Capp$^{1}$, Ranjit Ittyerah$^{1}$},\\
\textit{Elyse Migdal$^{1}$, Grace Choi$^{1}$, Emily Kopp$^{1}$, Bridget Loja$^{1}$, Eusha Hasan$^{1}$, Jiacheng Li$^{1}$, Karthik Prabhakaran$^{1}$},\\
\textit{Gabor Mizsei$^{1}$, Marianna Gabrielyan$^{1}$, Theresa Schuck$^{1}$, John Robinson$^{1}$, Daniel Ohm$^{1}$}, \\
\textit{Edward B. Lee$^{1}$, John Q. Trojanowski$^{1}$, Corey McMillan$^{1}$, Murray Grossman$^{1}$, David Irwin$^{1}$},\\
\textit{M. Dylan Tisdall$^{1}$, Sandhitsu R. Das$^{1}$, Laura E.M. Wisse$^{2}$, David A. Wolk$^{1}$, Paul A. Yushkevich$^{1}$}\end{tabular}}

\address{$^{1}$University of Pennsylvania, Philadelphia, USA $^{2}$Lund University, Lund, Sweden}

\begin{document}
\maketitle
\begin{abstract}
\textit{\textbf{Ex vivo}} MRI of the brain provides remarkable advantages over \textit{in vivo} MRI for visualizing and characterizing detailed neuroanatomy. However, automated cortical segmentation methods in \textit{ex vivo} MRI are not well developed, primarily due to limited availability of labeled datasets, and heterogeneity in scanner hardware and acquisition protocols. In this work, we present a high resolution 7~Tesla dataset of 32 \textit{ex vivo} human brain specimens. We benchmark the cortical mantle segmentation performance of nine neural network architectures, trained and evaluated using manually-segmented 3D patches sampled from specific cortical regions, and show excellent generalizing capabilities across whole brain hemispheres in different specimens, and also on unseen images acquired at different magnetic field strength and imaging sequences. Finally, we provide cortical thickness measurements across key regions in 3D \textit{ex vivo} human brain images. Our code and processed datasets are publicly available \href{https://github.com/Pulkit-Khandelwal/picsl-ex-vivo-segmentation}{\textbf{here}}.
\end{abstract}

\section{Introduction}
\label{sec:intro}

\textbf{\textit{Ex vivo}} MRI of the brain provides remarkable advantages over \textit{in vivo} MRI for visualizing detailed neuroanatomy and linking macroscopic morphometric measures such as cortical thickness to underlying cytoarchitecture and pathology \cite{mancini2020multimodal}. It helps in characterizing the underlying anatomy at the scale of subcortical layers \cite{augustinack2013medial}, such as hippocampal subfields in the medial temporal lobe (MTL) \cite{yushkevich2021three, ravikumar2020building}. Compared to \textit{in vivo} MRI, \textit{ex vivo} MRI is not affected by head or respiratory motion artifacts and has much less stringent time and specific absorption rate constraints. Compared to histology, it does not suffer from distortion or tearing of brain tissue, thereby giving flexibility in acquiring ultra-high resolution images. Indeed, \textit{ex vivo} MRI is often used to provide a 3D reference space onto which to map 2D histological images. Combined analysis of \textit{ex vivo} MRI and histology makes it possible to link morphological changes in the brain to underlying pathology as well as to generate anatomically correct parcellations of the brain based on cytoarchitecture \cite{schiffer2021convolutional}, and pathoarchitecture \cite{augustinack2013medial}.

There has been substantial work in brain MRI parcellation such as  \emph{FreeSurfer} \cite{fischl2012freesurfer} and recent efforts based on deep learning \cite{henschel2020fastsurfer, chen2018voxresnet}. However, these approaches focus on \textit{in vivo} MRI, and limited work has focused on developing automated segmentation methods for \textit{ex vivo} MRI segmentation. \textit{Ex vivo} segmentation methods have been region specific. Recent developments include automated deep learning methods for high resolution cytoarchitectonic mapping of the occipital lobe in 2D histological sections \cite{schiffer2021convolutional}. The work by \cite{iglesias2015computational} has developed an atlas to segment the MTL using manual segmentations in \textit{ex vivo} images. Yet, an \textit{ex vivo} segmentation method applicable to a variety of brain regions has yet to be described. Limited availability of \textit{ex vivo} 3D MRI segmentation algorithms may be explained by only a few groups focusing on whole brain \textit{ex vivo} image analysis and hence limited availability of specimens, scans, and labeled ground truth segmentations; greater heterogeneity in scanning protocols when compared to \textit{in vivo} structural MRI; larger image dimensions, greater textural complexity, and more profound imaging artifacts than in \textit{in vivo} MRI.

\begin{figure*}[t!]
  \includegraphics[width=\textwidth,height=8cm]{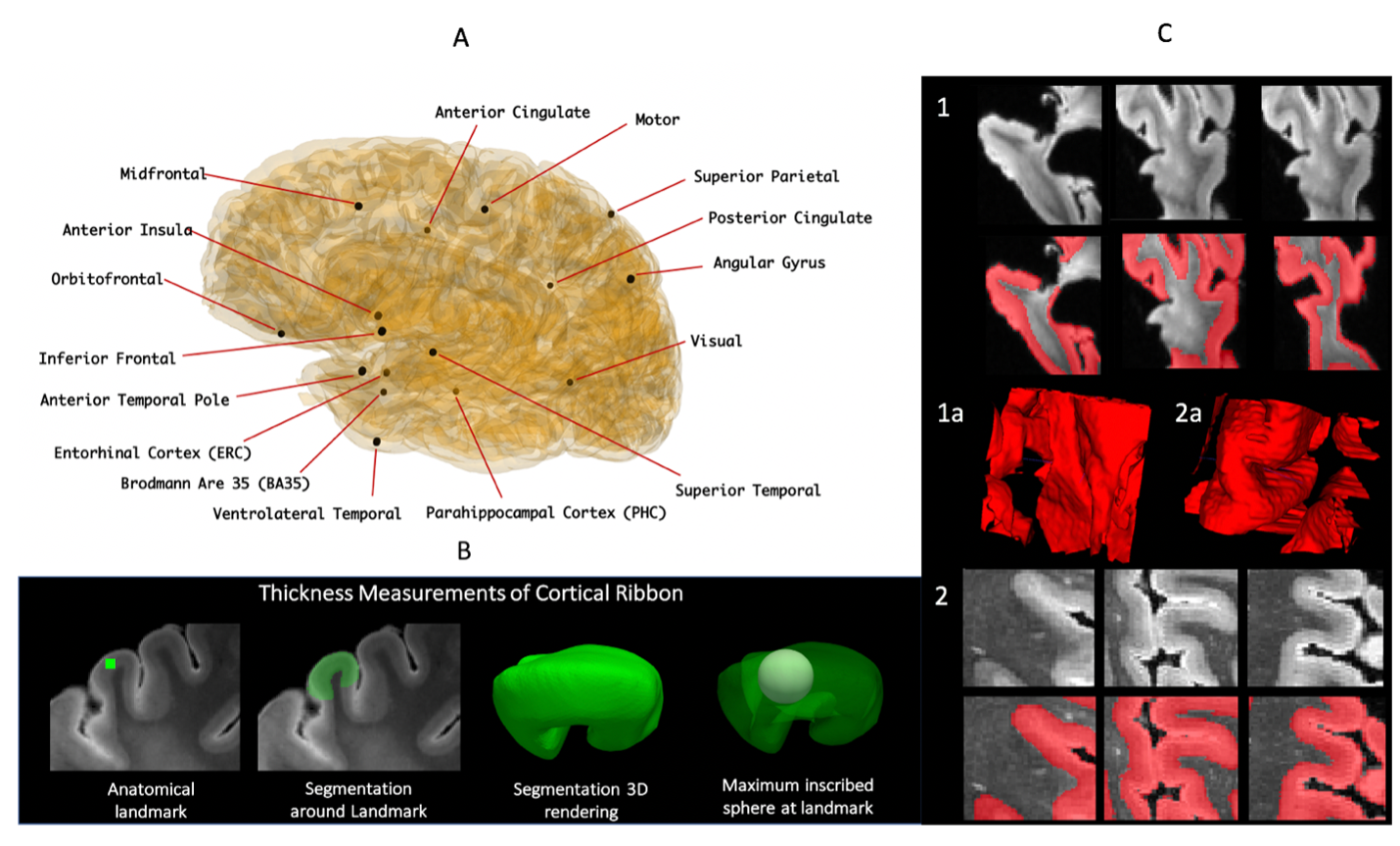}
  \caption{Thickness measurement pipeline and patch-level gray matter ground truth data. (A) Cortical thickness is measured at the 16 out of the 18 landmarks. \textit{Note that we do not measure thickness at Cornu Ammonis 1 (CA1) and subiculum as their segmentation requires Stratum Radiatum, Lacunosum, and Moleculare (SRLM) segmentation, which is currently not performed.} (B) A dot (shown: the motor cortex dot) is first placed to define an anatomical landmark, around which a semi-automatic level set segmentation of the surrounding cortical ribbon is provided. A maximally inscribed sphere is then computed using Voronoi skelentonization, and the diameter of the sphere gives thickness at that landmark. (C) Ground truth training data acquired by manually segmenting gray matter in 64 x 64 x 64 patches shown for two subjects (1 and 2) in three viewing planes with 3D renderings displayed in 1a and 2a respectively.}
  \label{fig:dots}
\end{figure*}

\textbf{In this work, we present a novel dataset of 32 high resolution (0.3 x 0.3 x 0.3 mm$^{3}$) 7~Tesla \textit{ex vivo} MRI scans of whole brain hemispheres of older adult patients with Alzheimer's Disease or Related Dementias (ADRD) or cognitively normal adults. We then benchmark nine deep learning neural architectures to segment cortical and sub-cortical gray matter in whole brain hemispheres, with limited patch-based training data.} We measure cortical thickness at several key locations in the cortex and correlate these automated measures with thickness measurements obtained using a user-guided semi-automated protocol. High consistency between these two sets of measures supports the use of deep learning based automated thickness measures for \textit{ex vivo} brain morphometry. Additionally, we show that networks trained on T2w images acquired at 7~Tesla are able to generalize to \textit{ex vivo} images obtained with T2*-weighted (T2*w) gradient echo images acquired at 7~Tesla, and \textit{ex vivo} images acquired at a lower field strength of 3~Tesla.

\begin{figure*}[!t]
  \includegraphics[width=\textwidth,height=6.5cm]{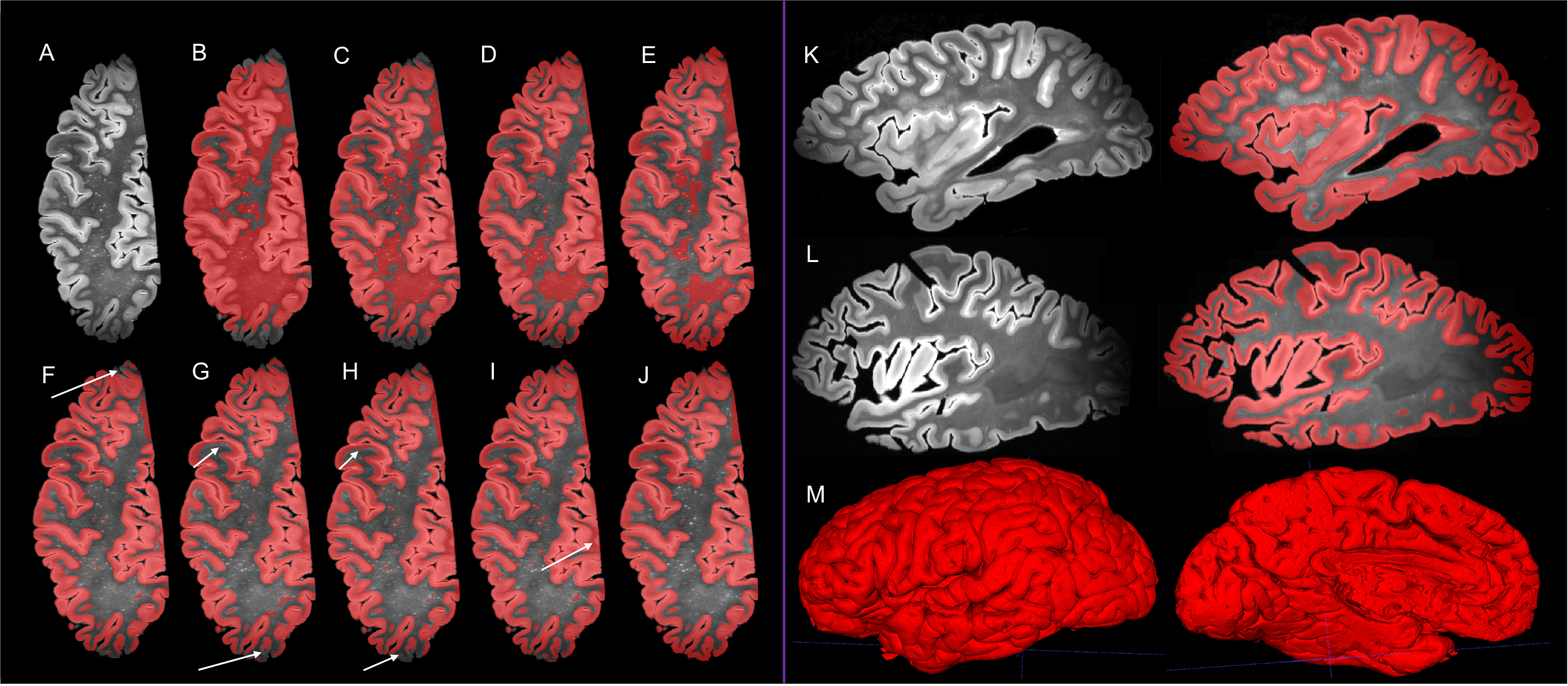}
  \caption{Cortical mantle segmentation across whole brain hemispheres of the different neural network architectures, on the 5th cross validation fold, for a given subject on the left panel. (A) an axial slice. (B) Attention Unet. (C) VNet. (D) VoxResNet. (E) 3D Unet. (F) AnatomyNet (CE + SE). (G) AnatomyNet (Vanilla). (H) AnatomyNet (CE). (I) AnatomyNet (SE). (J) nnU-Net. Right panel: (K-L) nnU-Net based segmentation shown on a sagittal plane for two subjects. (M) 3D rendering of the cortex from lateral and medial views for the subject shown in K.}
  \label{fig:all_results_compare}
\end{figure*}

\section{Materials}
\label{sec:materials}
\textbf{Image Acquisition.} We analyze a dataset of 32 \textit{ex vivo} whole-hemisphere MRI scans. Patients were selected for the study from our ongoing research autopsy program. Data was drawn from 11 females (Age: 64-94) and 21 males (Age: 54-97) with Alzheimer's Disease or Related Dementias (ADRD) or cognitively normal adults. Human brain specimens were obtained in accordance with the University of Pennsylvania Institutional Review Board guidelines. Specimens were scanned after atleast a 4 week fixation period. T2w images were acquired using a 3D-encoded T2 SPACE sequence with 0.28 mm isotropic resolution, 3 s repetition time (TR), echo time (TE) 383 ms, turbo factor 188, echo train duration 951 ms, bandwidth 348 Hz/px. All data was acquired on a Siemens MAGNETOM Terra 7~Tesla scanner using a custom birdcage transmit/receive coil. Sample slices are shown in Fig. \ref{fig:all_results_compare}.

\textbf{Patch-level Gray Matter Segmentation.} To train the neural networks, we sampled five 3D image patches of size 64 x 64 x 64 around the orbitofrontal, anterior temporal, inferior prefrontal, primary motor, and primary somatosensory cortices from 6 brain hemispheres, resulting in a total of 30 patches. Fig. \ref{fig:dots} C shows sample patch images and the corresponding ground truth labels with 3D renderings. Five manual raters, divided into groups of two and three, labeled gray matter as the foreground, and rest of the image as the background using a combination of manual tracing and the semi-automated segmentation tool in ITK-SNAP software \cite{yushkevich2019user}. Inter-rater reliability scores were computed for these manual segmentations in terms of Dice Coefficient (DSC): Raters 1\&2: 95.26 $\pm$ 1.37 \%, Raters 1\&3: 94.64 $\pm$ 1.64 \%, Raters 2\&3: 94.54 $\pm$ 1.20 \%, Raters 4\&5: 92.04 $\pm$ 4.26 \%.

\begin{figure*}[!t]
    \centering
  \includegraphics[scale=1]{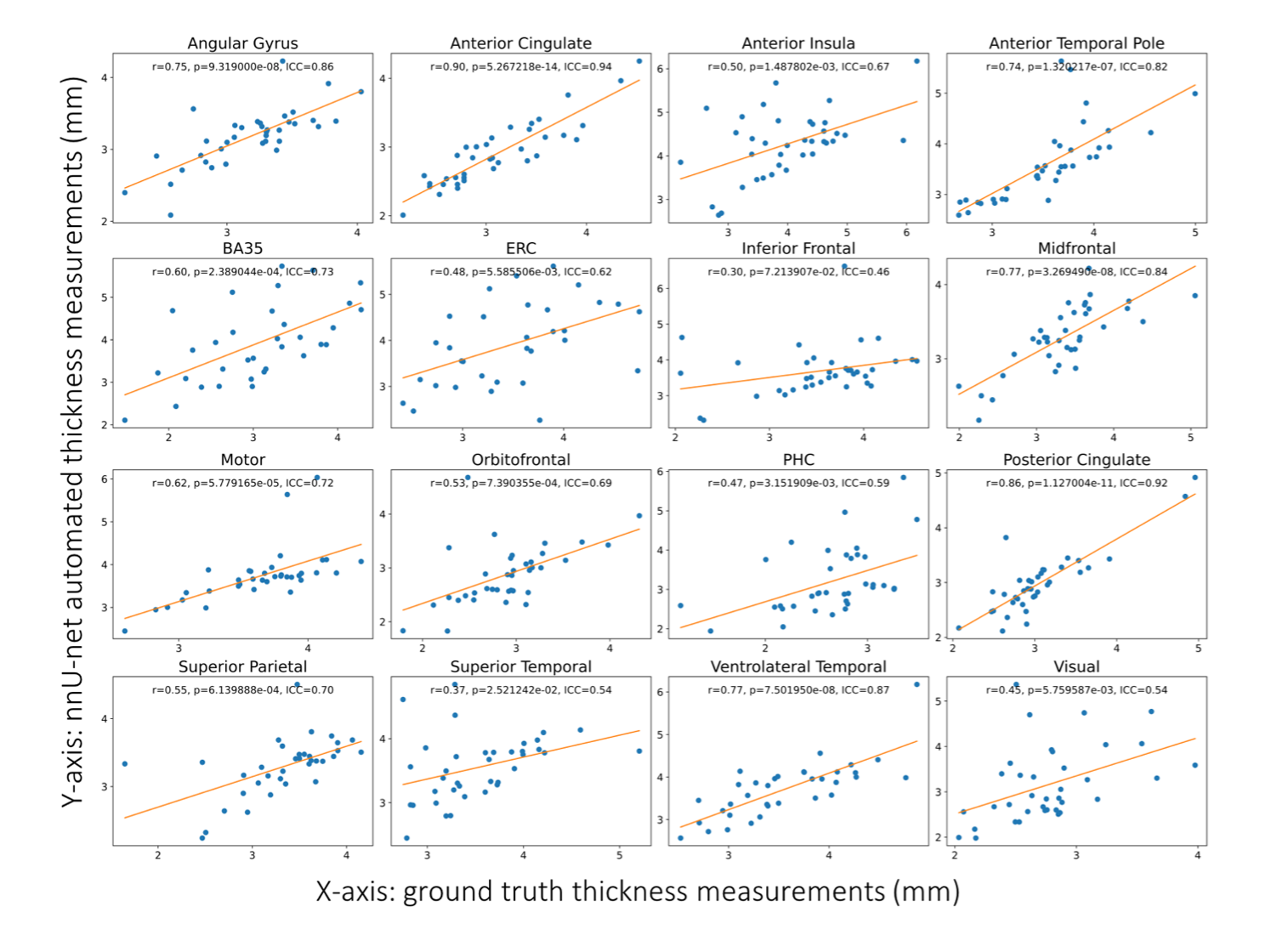}
  \caption{Cortical thickness measurements at the 16 cortical regions. Shown are correlation plots between cortical thickness measured by automated nnU-Net (y-axis) and ground truth manual segmentations (x-axis). Within each plot, we tabulate Pearson's correlation coefficient (r), p-value, and the Average fixed raters Intra-class correlation coefficient (ICC) scores.}
  \label{fig:graphs}
\end{figure*}

\textbf{Thickness measurements at key cortical locations.}
To obtain localized quantitative signatures of cortical morphometry at our Center, in each of the 32 hemispheres, we identified 18 cortical landmarks including 5 in the medial temporal lobe (Fig. \ref{fig:dots} A): visual, midfrontal, orbitofrontal, motor, anterior and posterior cingulate, superior and ventrolateral temporal, anterior temporal pole, anterior insula, inferior frontal, angular gyrus, superior parietal, Entorhinal Cortex (ERC), Brodman Area 35 (BA35), Parahippocampal Cortex (PHC), Cornu Ammonis 1 (CA1) and subiculum. These locations were chosen to gather neuropathology data which is part of a separate ongoing research project. To measure cortical thickness at these locations, we use the pipeline developed in \cite{wisse2021downstream}, shown in Fig. \ref{fig:dots} B.

\section{Methods}
\label{sec:methods}
We benchmark variants of popular biomedical image segmentation deep learning models: (1.) nnU-Net \cite{isensee2021nnu}; four variants of AnatomyNet \cite{zhu2019anatomynet} based on squeeze-and-excitation blocks \cite{rickmann2019project}: (2.) Spatial excitation AnatomyNet (SE), (3.) Channel excitation AnatomyNet (CE), (4.) AnatomyNet (Vanilla) \cite{zhu2019anatomynet}, (5.) Channel-spatial excitation AnatomyNet (CE + SE); (6.) 3D Unet-like network \cite{khandelwal2020domain}; (7.) VoxResNet \cite{chen2018voxresnet}; (8.) VNet \cite{milletari2016v} ; and (9.) Attention Unet \cite{oktay2018attention}. We use PyTorch 1.5.1 and Nvidia Quadro RTX 5000 GPUs to train the models using user-annotated patches described above. Patches were standardized, and then normalized between 0 and 1. We implemented all the network architectures within the nnU-Net framework \cite{isensee2021nnu} and thus systematically evaluated the nine deep learning architectures in a five-fold cross-validation experiment under matched conditions (i.e., same split of data into training/validation/testing subsets; same data augmentation strategy, same hyper-parameter tuning strategy).

\textbf{Evaluation.} First, we compare the performance of the deep learning architectures at patch-level by reporting Dice Coefficient (DSC) and Hausdorff Distance 95th percentile (HD95) in a five-fold cross-validation setting for the 30 patches. We then employ the best performing model, based on qualitative results as shown in Fig. \ref{fig:all_results_compare}, to segment the cortical mantle in whole hemispheres. We compute the thickness of the cortical mantle around 13 landmarks as described in Section \ref{sec:materials}. We then correlate the cortical thickness of manual, and automated segmentations via Pearson's correlation coefficient with t-distribution as the test statistic reporting the p-value (with 0.05 significance level), and the Average fixed raters Intra-class Correlation Coefficient (ICC) for the 13 cortical locations.

\section{Results and Discussion}
\label{sec:results}

\subsection{Deep learning segmentations}
\label{sec:results_DL}
Table 1 tabulates the performance of different networks across 6-fold cross validation. In terms of DSC, AnatomyNet and its variants performs at par with VoxResNet, and outperforms the rest of the networks. But, superior DSC performance does not translate equally well to whole hemisphere segmentations as evident by qualitative results in Fig. \ref{fig:all_results_compare}. We observe that inferior performing models (\ref{fig:all_results_compare} B-E) mislabel white matter as gray matter. The AnatomyNet and its variants (\ref{fig:all_results_compare} F-I) are able to distinguish gray matter from white matter, but fail to segment the low intensity anterior and posterior regions, virtue of coil limitations, shown in white arrows. There are also some under-segmentations of the cortex (white arrows). Fig. \ref{fig:all_results_compare} J-L depicts that the best performing model is nnU-Net, which clearly demarcates GM/WM boundary, segments regions with low signal, which were \textit{not} included in the training patches, making the performance of nnU-Net even more remarkable. Fig. \ref{fig:all_results_compare} M shows 3D renderings of whole cortical segmentation. Hence, we use nnU-Net to segment the cortical mantle in whole hemisphere across the 32 subjects.

\begin{table}[t]
    \label{table:results}
    \caption{Five-fold cross validation Dice Coefficient (DSC) and Hausdorff Distance 95th percentile (HD95) scores between ground truth and automated patch-level segmentations.}
    \centering
        {\begin{tabular}{c|c|c}
    \makecell{\textbf{Deep learning method}} & \makecell{\textbf{DSC} \textbf{(\%)}} & \makecell{\textbf{HD95} \textbf{(mm)}} \\
    \hline
    nnU-Net &
    93.98 $\pm$ 5.25 & 
    0.49 $\pm$ 0.45 \\
    \hline
    AnatomyNet (SE) & 
    94.84 $\pm$ 3.84 & 
    0.45 $\pm$ 0.42 \\
    \hline
    AnatomyNet (CE) & 
    94.91 $\pm$ 3.27 & 
    0.45 $\pm$ 0.42 \\
    \hline
    AnatomyNet (Vanilla) & 
    94.86 $\pm$ 3.83 & 
    0.46 $\pm$ 0.44 \\
    \hline
    AnatomyNet (CE + SE) & 
    94.66 $\pm$ 3.79 & 
    0.47 $\pm$ 0.44 \\
    \hline
    3D Unet & 
    93.57 $\pm$ 5.22 & 
    0.58 $\pm$ 0.51 \\
    \hline
    VoxResNet & 
    94.84 $\pm$ 4.00 & 
    0.45 $\pm$ 0.42 \\    
    \hline
    VNet & 
    90.84 $\pm$ 5.93 & 
    0.99 $\pm$ 0.56 \\
    \hline
    Attention Unet & 
    93.65 $\pm$ 4.91 & 
    0.62 $\pm$ 0.66
  \end{tabular}}
\end{table}

\subsection{Cortical Thickness Measurements}
\label{sec:results_cortical}
We correlate thickness (mm) between ground truth and automated nnU-Net based segmentations at 16 out of the 18 landmarks. Note that we do not measure thickness at Cornu Ammonis 1 (CA1) and subiculum as their segmentation requires Stratum Radiatum, Lacunosum, and Moleculare (SRLM) segmentation, which is currently not performed. Fig. \ref{fig:graphs} shows good agreement between ground truth and automated thickness with 8 regions having correlation coefficient (r) greater than 0.6 with 15 regions reaching statistical significance at p $<$ 0.05, except in inferior frontal region. We also observe high ICC scores with 9 regions having ICC greater than 0.7, which confirms that automated segmentations are accurate to give desirable cortical thickness measurements.

\begin{figure*}[!t]
    \centering
  \includegraphics[scale=0.6]{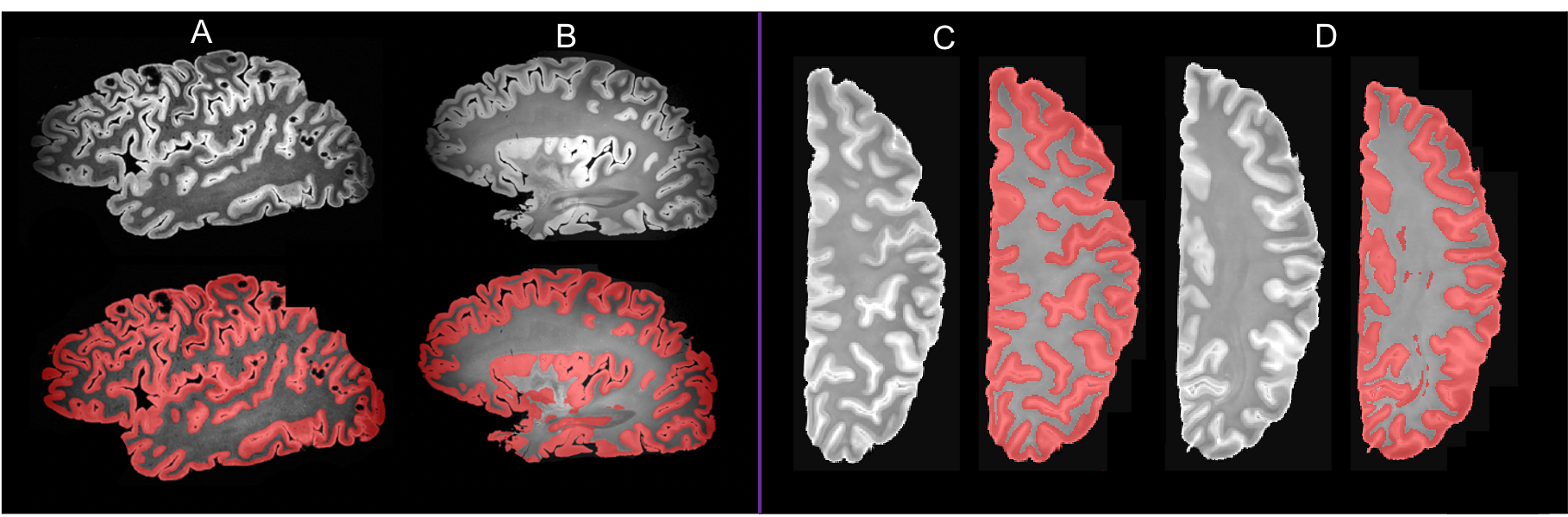}
    \caption{Generalization to other unseen imaging sequences. (A-B) image acquired using T2*w gradient echo, (C-D) 3~Tesla image at lower resolution of 0.5 x 0.5 x 0.5 mm$^{3}$.}
\label{fig:results_other_data}
\end{figure*}

\subsection{Generalization to other imaging sequences}
\label{sec:results_generalize}
In Fig. \ref{fig:results_other_data}, we qualitatively show that our model trained on 7~Tesla 0.3 x 0.3 x 0.3 mm${^3}$ T2w images, is able to generalize well to MRI sequences and resolutions unseen during training. Segmentation results on T2*w gradient echo \textit{ex vivo} images, acquired \cite{tisdall2021joint} at 0.28 mm and 0.16 mm isotropic resolution, are shown in Fig. \ref{fig:results_other_data} A and B respectively. Fig. \ref{fig:results_other_data} C and D show that our model is able to segment gray matter in the publicly available \textit{ex vivo} T2w image acquired at 3~Tesla at a lower resolution of 0.5 x 0.5 x 0.5 mm$^{3}$ \cite{mancini2020multimodal}.

\section{Conclusion and Future Work}
\label{sec:future_work}
Our results show that even using limited patch-level training data from six subjects, nnU-Net (and to a lesser extent AnatomyNet) is able to generate high-quality segmentations of cortical and subcortical gray matter in \textit{ex vivo} MRI of brain hemispheres, generalizing well to areas of low contrast unseen during training, as well as to other MRI protocols, field strengths, and resolutions. Moreover, thickness measures derived from nnU-Net segmentations concur with user-supervised thickness measurements, suggesting the feasibility of fully automated cortical thickness analysis in \textit{ex vivo} MRI analogous to the way \emph{FreeSurfer} is used for \textit{in vivo} MRI morphometry. A limitation of our approach is that it does not separate subcortical gray matter from the cortex. In future work, we intend to address this limitation using anatomical priors, and develop techniques for groupwise normalization of \textit{ex vivo} MRI and correlate cortical thickness with neuropathology.


\vfill
\pagebreak

\bibliographystyle{IEEEbib}
\bibliography{refs}

\end{document}